# Persistent Spin Helix Manipulation by Optical Doping of a CdTe Quantum Well

F. Passmann,[1] S. Anghel,[1] T. Tischler,[1] A. V. Poshakinskiy,[2] S. A. Tarasenko,[2] G. Karczewski,[3] T. Wojtowicz,[3,4] A. D. Bristow[1,5] and M. Betz[1]

[1] *Experimentelle Physik 2, Technische Universität Dortmund, Otto-Hahn-Straße 4a, D-44227 Dortmund, Germany*
[2] *Ioffe Institute, St. Petersburg 194021, Russia*
[3] *Institute of Physics, Polish Academy of Science, Aleja Lotników 32/46, PL-02668 Warsaw, Poland*
[4] *International Research Centre MagTop, Aleja Lotników 32/46, PL-02668 Warsaw, Poland*
[5] *Department of Physics and Astronomy, West Virginia University, Morgantown, WV 26506-6315, U.S.A.*
*E-mail address: markus.betz@tu-dortmund.de*

Time-resolved Kerr-rotation microscopy explores the influence of optical doping on the persistent spin helix in a [001]-grown CdTe quantum well at cryogenic temperatures. Electron spin diffusion dynamics reveal a momentum-dependent effective magnetic field providing SU(2) spin-rotation symmetry, consistent with kinetic theory. The Dresselhaus and Rashba spin-orbit coupling parameters are extracted independently from rotating the spin helix with external magnetic fields applied parallel and perpendicular to the effective magnetic field. Most importantly, a non-uniform spatiotemporal precession pattern is observed. The kinetic theory framework of spin diffusion allows for modeling of this finding by incorporating the photocarrier density into the Rashba ($\alpha$) and the Dresselhaus ($\beta_3$) parameters. Corresponding calculations are further validated by an excitation-density dependent measurement. This work shows universality of the persistent spin helix by its observation in a II-VI compound and the ability to fine-tune it by optical doping.

Spin-orbit (SO) interaction in two-dimensional electron gases (2DEGs) is responsible for a broad range of phenomena, including spin Hall effects [1–3] and spin textures [4,5], such as the persistent spin helix (PSH). A PSH occurs when parameters associated with the bulk Dresselhaus [6] and structural Rashba [7] inversion asymmetries are equal in strength [8–10]. This results in a momentum-dependent effective magnetic field, $\boldsymbol{B}_{\mathrm{SO}}(\boldsymbol{k})$, providing the SU(2) symmetry in which the 2DEG exhibits a unidirectional spin grating (or helical spin-density wave) [11]. The reduced symmetry suppresses the spin-orbit dephasing [12], increasing the spin lifetime by several orders of magnitude [10,13].

PSH texturing shows promise for spintronic applications, because Dresselhaus ($\beta$) and Rashba ($\alpha$) SO coupling can be readily tailored by material choice and device design [14]. For example, 2DEGs in *zincblende* nanostructures, such as [001]-grown GaAs quantum wells (QWs) with modulation doping, are naturally suited to balance $\alpha$ and $\beta$ independently, adjustable through doping and well width [11]. External magnetic fields vectorially add to $\boldsymbol{B}_{\mathrm{SO}}$, allowing for the determination of $\alpha$ and $\beta$ [15]. In addition, electric fields induced by a back gate provide direct electrical control over $\boldsymbol{B}_{\mathrm{SO}}$ and the spin properties of the system [16–18]. Dual modulation-doping geometries lead to subbands either side of a soft barrier within the well [19,20] demonstrating a stretchable PSH by tuning $\alpha$ and $\beta$ together [21]. Moderate in-plane electric fields will drive a spin helix to propagate [22], while the drift velocity can modify the PSH wavevector [23,24].

Spin control schemes based on optical doping of lightly *n*-doped GaAs QWs with photocarriers have been explored [25,26], indicating that PSH dynamics could also be manipulated in this way. Hence, in this letter, the effect of optical doping on the spatiotemporal evolution of spin

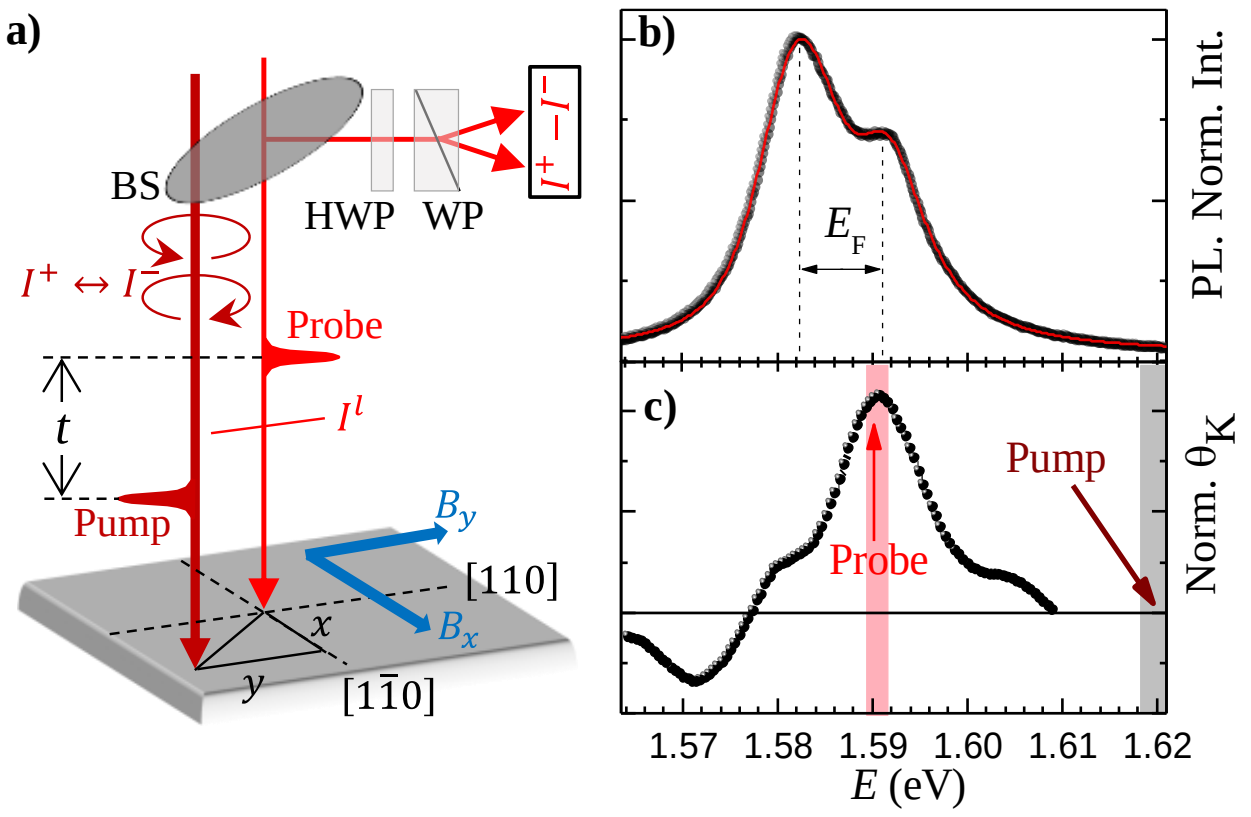

**Fig. 1 (a)** Schematic diagram of the experimental setup. Pump pulses are modulated between left and right circular polarization and moved in the *xy*-plane, defined by the crystal symmetry. Probe pulses are linearly polarized and Kerr rotation is obtained using a half-wave plate (HWP) and Wollaston prism (WP). **(b)** Photoluminescence spectrum revealing the Fermi energy $E_F$. **(c)** A normalized Kerr-rotation spectrum at $t = 0$ ps, with pump and probe spectral positions illustrated for all following experiments polarization $S_z(x,y,t)$ is explored in a CdTe 2DEG. CdTe has a larger SO coupling, and accordingly shorter spin lifetimes, and a larger g-factor (absolute) than GaAs [27], in addition to generally lower diffusion coefficients. While the observation of the PSH is therefore more challenging in CdTe than in GaAs 2DEGs, the phenomenon is not limited to the latter system – confirmed by its observation at a strained $LaAlO_3/SrTiO_3$ interface [28]. Here, experiments are performed in a regime where optical doping is sufficient to modify the diffusive expansion of photocarrier, their spin dynamics and the wavelength of the spin helix.

The sample consists of a 200-Å CdTe QW with $Cd_{0.74}Mg_{0.26}Te$ barriers, grown by molecular beam epitaxy on a [001]-GaAs substrate [29]. Iodine modulation doping produces a 2DEG concentration of $n_0 \approx 3.4\times10^{11}$ cm$^{-2}$ and a mobility of $\mu \approx 4.2\times10^5$ cm$^2$/Vs, determined from (Hall and conductivity) magneto-transport measurements at ~4.0 K. Figure 1(b) shows the photoluminescence (PL) spectrum for illumination with a HeNe laser. Peaks centered at 1.582 eV and 1.591 eV correspond to recombination from conduction-band minimum and Fermi edge respectively [30]. The spectral separation is the 2DEG Fermi energy $E_F$ = 9.0±0.1 meV, corresponding to $n_0$ = (3.5±0.1)×10$^{11}$ cm$^{-2}$, in good agreement with transport data.

For the [001]-grown *zincblende* QWs, the balance of Dresselhaus and Rashba SO coupling leads to SU(2) spin rotation symmetry in the $\boldsymbol{k}$-dependent effective magnetic field

$$\boldsymbol{B}_{\text{SO}} = \frac{2}{g\mu_B}\begin{pmatrix}[\alpha+\beta]\cdot k_y\\ [\beta-\alpha]\cdot k_x\end{pmatrix}, \quad (1)$$

where $g$ is the effective Landé $g$-factor, $\mu_B$ is the Bohr magneton, $k_x$ and $k_y$ are the in-plane wavevectors with $x \parallel [1\bar{1}0]$ and $y \parallel [110]$, and $\alpha = \gamma_R E_z$ is related to the effective electric field $E_z$ in the QW by the Rashba coupling constant $\gamma_R$. Dresselhaus SO coupling can be described by $\beta = \beta_1 - \beta_3$, where $\beta_1 = \gamma_D\langle k_z^2\rangle$ depends on the bulk Dresselhaus coupling constant $\gamma_D$ [31] and the wavevector component $k_z$ due to quantum confinement in the QW, and $\beta_3 = \gamma_D k_F^2/4$ depends on the Fermi wavevector $k_F = \sqrt{2\pi n_0}$. For the well width and 2DEG carrier density in our sample, it is estimated that $\beta_1 \approx 4.0\beta_3$. Spin polarization of the photoexcited carriers evolves due to carrier diffusion in the presence of $\boldsymbol{B}_{\text{SO}}$, which can be directly measured using time-resolved polar magneto-optic Kerr microscopy [24,32].

Figure 1(a) shows the experimental scheme, wherein shaped pulses from a 60-MHz modelocked Ti:sapphire oscillator are used to photoexcite spin-polarized electrons in the conduction band at a photon energy above $E_F$. $S_z(x,y,t)$ is mapped using a second linearly-polarized pulse and differential detection captures the Kerr rotation signal $\theta_K$. The pump and probe can be independently tuned, as illustrated by the normalized $\theta_K$ spectrum in Fig. 1(c). All following measurements are performed with the probe tuned to the peak Kerr signal at 1.59 eV. Pump and probe pulse peak intensities are set to 3.53 MW/cm$^2$ and 2.36 MW/cm$^2$ respectively. Both pulses impinge the sample through a 50× objective giving FWHM spot sizes of ~3(1) µm for the pump(probe), with control of the relative pump-probe overlap by the microscope design [33]. The delay time between pump and probe pulses, $t$, can be varied over a range of ~1.8 ns, with a pulse-shaped temporal resolution of <1 ps. A Voigt-geometry ≤235 mT magnetic field can be externally applied.

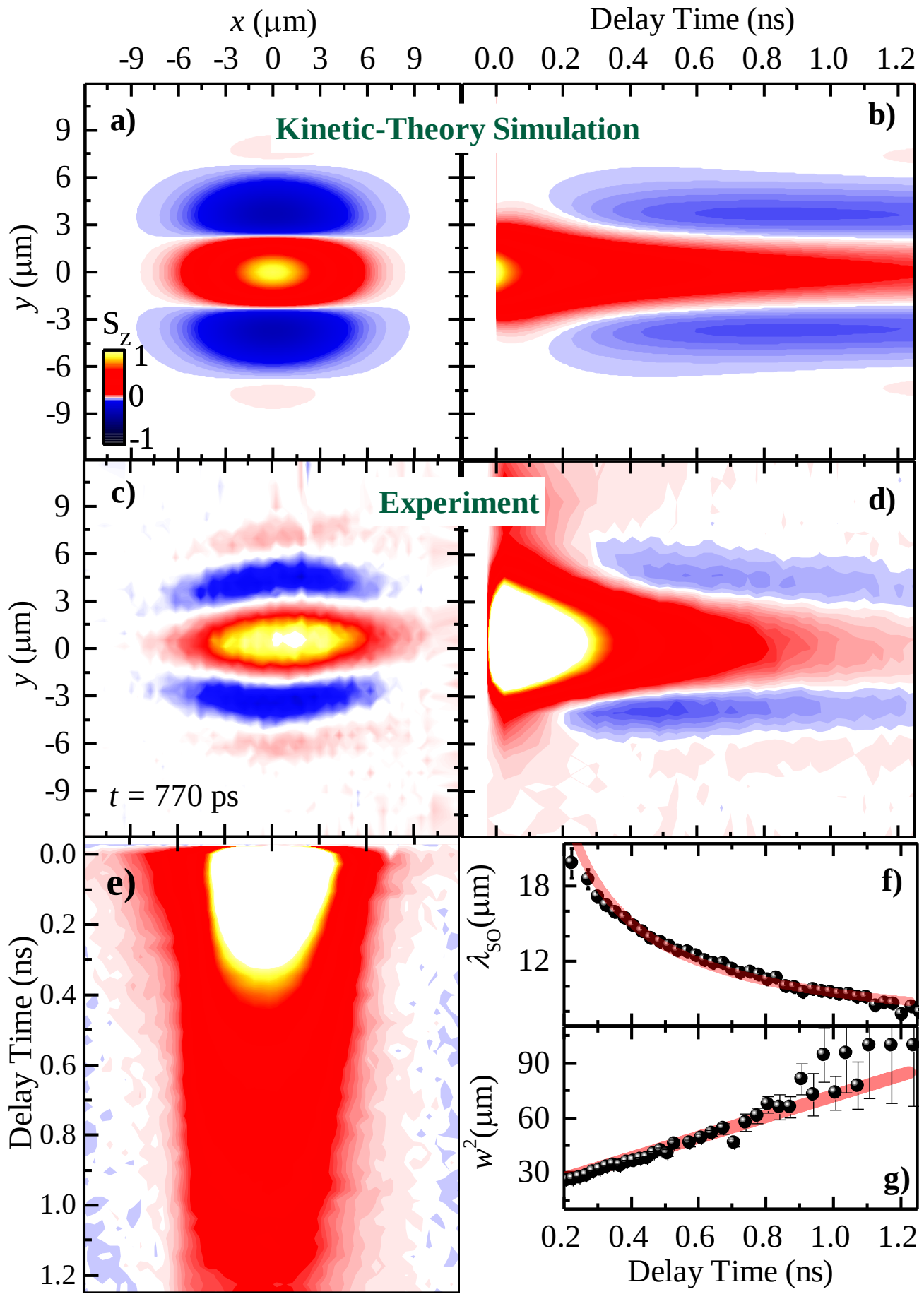

**Fig. 2** Time-resolved Kerr-rotation microscopy of a CdTe 2DEG. **(a,b)** Kinetic theory simulations and **(c-e)** experimental measurements of $S_z(x,y,t)$, where in each panel one of the three space or time axes is fixed. Extracted **(f)** spin precession length $\lambda_{SO}(t)$ and **(g)** squared Gaussian width $w^2(t)$ of the experimental data, from which the diffusion coefficient can be determined.

Figure 2 (c)-(e) show $S_z(x,y,770\text{ ps})$, $S_z(0,y,t)$ and $S_z(x,0,t)$ respectively, experimentally mapping diffusion of the spin-polarized photocarriers as they approach the PSH mode. This spin diffusion reveals the predominant orientation of $\boldsymbol{B}_{\text{SO}}$, through Larmor precession in the $y$ direction, but not in the $x$ direction. The spatial profile is well described by

$$S_z(x,y) = A\cdot e^{-\frac{4\ln(2)(x^2+y^2)}{w^2}}\cdot\cos\left[\frac{2\pi(y-y_1)}{\lambda_{\text{SO}}}\right], \quad (2)$$

where $w$ represents the FWHM of the Gaussian envelope. Due to the finite spot size and the carrier diffusion, the envelope grows in time according to $w^2(t) = w_0^2 + 16\ln(2)\,D_s t$, where $D_s$ is the spin-diffusion coefficient and $w_0$ is the initial FWHM defined by the laser spot. The cosine term assumes a PSH, captures the spin texture centered at $y_1$ and characterizes the distance over which it takes for one complete spin precession to occur, known as the spin precession length,

$$\lambda_{\mathrm{SO}} = \lambda_0\{1 + w_0^2/[16\ln(2)D_s t]\}, \quad (3)$$

where $\lambda_0 = \pi\hbar^2/(m^*|\alpha + \beta|)$ is the PSH spin precession length [34,35], $m^* = 0.094\, m_e$.

Figure 2(f) and (g) show $\lambda_{\mathrm{SO}}(t)$ and $w^2(t)$, determined by fitting $S_z(0,y,t)$ and $S_z(x,0,t)$ with Eq.(2) for fixed delay times. Fit lines show agreement with the above analytical expressions. Namely, $\lambda_{\mathrm{SO}}$ decays in time towards the PSH spin precession length, $\lambda_0$ = 5.6±0.1 μm, and $w^2$ increases linearly with time due to diffusion, from which it is found that $D_s$ = 50±10 cm$^2$/s with $w_0$=3.3±0.2 μm.

Applying two orthogonal, in-plane, external magnetic fields shifts or rotates the spin helix depending on the field orientation, from which $\alpha$ and $\beta$ can be extracted [15]. Figure 3(a) and (b) show $S_z(0,y,t)$ and $S_z(x,0,t)$ for the applied fields $|B_x|$ = $|B_y|$ = 231 mT. The external magnetic field adds vectorially to $\boldsymbol{B}_{\mathrm{SO}}$, such that electrons with a specific momentum, obeying $\hbar k_y = m^*\, dy_1/dt$ or $\hbar k_x = m^*\, dx_1/dt$ (where $x_1$ is the $x$-direction equivalent of $y_1$) do not undergo Larmor precession. These conditions lead to the observed stripe patterns in $S_z(0,y,t)$ and $S_z(x,0,t)$, which can be described by $dy_1(B_x)/dt = qB_x/(\alpha + \beta)$ and $dx_1(B_y)/dt = qB_y/(\beta - \alpha)$, where $q = \hbar g\mu_{\mathrm{B}}/2m^*$. These formulae are used to analyze the data for $t > 1$ ns, when the PSH mode is formed, giving $\alpha + \beta = 4.08\pm0.33$ meVÅ and $\alpha - \beta =$ 0.63±0.01 meVÅ. From these values, it is found that $\alpha =$ 2.35±0.17 meVÅ, $\beta =$ 1.73±0.16 meVÅ. For a 20-nm-wide QW this yields $\gamma_{\mathrm{D}}$ = 9.10±0.84 eVÅ$^3$, in good agreement with literature [31].

The experimental results are modeled within the framework of a kinetic equation for the complete spin distribution $\boldsymbol{S}$ described elsewhere [24,35] by

$$\frac{\partial \boldsymbol{S}}{\partial t} = D_s\frac{\partial^2 \boldsymbol{S}}{\partial r^2} - \boldsymbol{\Gamma}\boldsymbol{S} - \left(\boldsymbol{\Lambda}\frac{\partial}{\partial r}\right) \times \boldsymbol{S} + \boldsymbol{\Omega}_L \times \boldsymbol{S} \quad (4)$$

where $\boldsymbol{r} = (x, y)$, $\boldsymbol{\Gamma}$ is the Dyakonov-Perel spin-relaxation-rate tensor with diagonal components $\Gamma_{xx} = D_s[2m^*(\alpha - \beta)/\hbar^2]^2$, $\Gamma_{yy} = D_s[2m^*(\alpha + \beta)/\hbar^2]^2$, $\Gamma_{zz} = \Gamma_{xx} + \Gamma_{yy}$, $\boldsymbol{\Lambda}$ is the tensor describing the spin precession during diffusion with nonzero components $\Lambda_{xy} = 4D_s m^*(\alpha + \beta)/\hbar^2$, $\Lambda_{yx} = -4D_s m^*(\alpha - \beta)/\hbar^2$ and $\boldsymbol{\Omega}_L$ is the Larmor precession frequency due to applied magnetic fields. Using the input values of $w_0$, $D_s$, $\alpha$ and $\beta$ extracted from the measurements, Fig. 2(a) and (b) show kinetic-theory simulations of $S_z(x,y,770\ \mathrm{ps})$ and $S_z(0,y,t)$ that reproduce the spatial growth of the spin-polarization signal with increasing time as it relaxes towards the PSH mode, using the experimental spin lifetime of ~300 ps.

Figure 3 (c) and (d) show results of kinetic-theory calculations [24,35] using input parameters determined from the experiment and including precession in the applied magnetic field with the electron $g$-factor ($g$ = 1.8) determined independently. The calculations reproduce the stripe pattern observed in the experimental data, confirming the general spin dynamics [34]. Comparing the calculations with experiment highlights a saddle-point-like (S-shape) stripe behavior in the experimental $S_z(0,y,t)$; see the dotted guide to the eye in Fig.3(a) which is not replicated in the dotted guide placed at a similar location in Fig. 3(c). Although the S-shape behavior is striking in $S_z(0,y,t)$, it is also present in $S_z(x,0,t)$. The slope of each stripe shows a spatial variation that is different in the center from the edges, as if $dy_1(y)/dt$ is modified by the initial optical excitation.

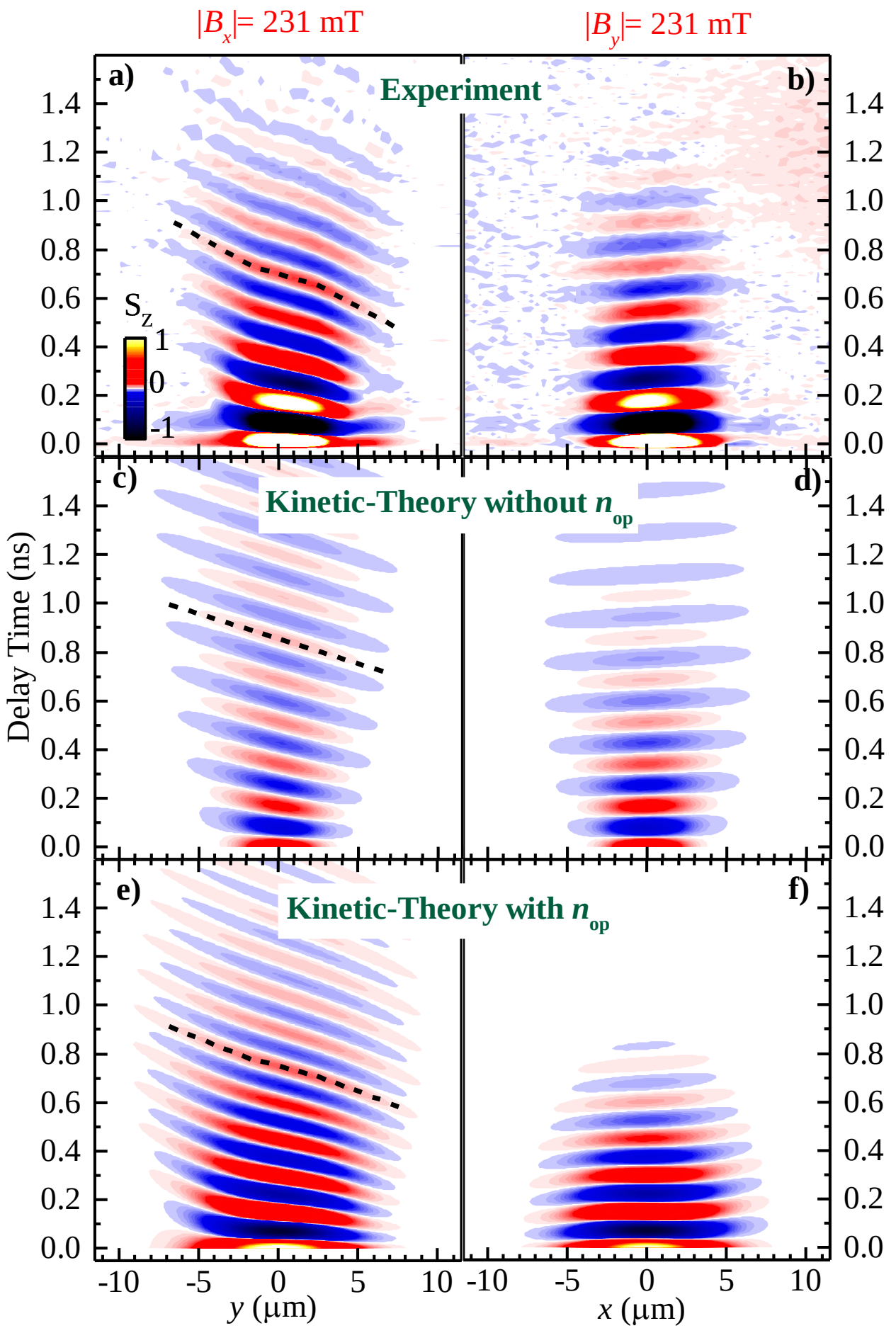

**Fig. 3** Spin polarization $S_z(x,y,t)$ in the presence of $B_x$ (left) and $B_y$ (right): **(a,b)** experimental, **(c,d)** kinetic theory simulation and **(e,f)** phenomenological model of carrier density and spin diffusion coefficients dependent $\lambda_{SO}$.

Considering optical excitation of the QW, a dynamic non-equilibrium carrier concentration is produced with a Gaussian spatial distribution. Consequently, the time and spatial dependence of the electron density after a focused pump pulse can be described by

$$n_{\mathrm{e}} = n_0 + n_{\mathrm{op}}\ln(2)\,\pi\left[\frac{w_0}{w(t)}\right]^2 e^{-4\ln(2)[x^2+y^2]/w^2}\, e^{-t/t_c}, \quad (5)$$

where $n_{\text{op}}$ is the average initial photocarrier density within the laser spot, $t_c \sim 1ns$ is the carrier lifetime determined from transient reflection. The local increase in the electron density leads to a local increase in the density-dependent Dresselhaus parameter $\beta_3$ which can be rewritten as

$$\beta_3(x,y,t) = \gamma_{\text{D}}\pi\, n_{\text{e}}(x,y,t)/2. \qquad (6)$$

Optical doping modifies also the Rashba spin-orbit coupling parameter $\alpha$ since the photocarriers screen the electric field $E_{\text{z}}$. The depolarization field $\delta E_{\text{z}}$ produced by a small carrier density $\delta n_{\text{e}}$ can be estimated as $\delta E_{\text{z}} = -\rho(e^2/\varepsilon)(m^* l^3/\hbar)E_z \delta n_{\text{e}}$, where $\varepsilon$ is the dielectric constant, $l$ is the QW width, and $\rho \approx 0.3$ is a dimensionless parameter which is defined by the QW design and has only a weak dependence on the carrier density. This equation yields

$$\alpha(x,y,t) = \alpha_0 e^{-\rho m^* e^2 l^3/(\varepsilon\hbar)\,(n_{\text{e}}(x,y,t)-n_0)}, \qquad (7)$$

where $\alpha_0$ is the as-grown Rashba parameter. Thus, increasing the optical-excited carrier density increases $\beta_3$ and reduces $\alpha$ and $\beta$. It follows from Eq. (3), $\lambda_0 = \pi\hbar^2/(m^*|\alpha+\beta_1-\beta_3|)$, that the spin precession takes place now in a spatially varying effective magnetic field and the spin precession length $\lambda_0$ is larger towards the center of the laser spot. Finally, the spin diffusion coefficient in the $x$ direction depends on the photocarrier density through the Fermi distribution.

As a consequence, because $\alpha$, $\beta_3$ and $D_{\text{s}}$ are dependent on optical doping, then $S_z(x,y,t)$ must become spatially modified by $n_{\text{e}}(x,y,t)$. Validation of this prediction requires comparison of intensity-dependent experiments to calculations where the spatial dependence is considered. Figure 4(a) shows normalized simulations of $S_z(y,0.75\text{ns})$ based on Eq (4), where $n_{\text{op}}$ varies $\boldsymbol{\Gamma}$ and $\boldsymbol{\Lambda}$ over a range to match the supporting experiments and with no additional Larmor precession from an external magnetic field. Only positive $y$ values of the data are shown, because the evolution is symmetric without an applied $B$-field. The model shows that for an increase in excitation density there is an increase in the modulation depth and spin precession length, as typified by the trends illustrated with arrows. For comparison, Fig. 4(b) shows normalized experimental $S_z(0,y,0.75\text{ns})$ recorded without an applied magnetic field and for a range of optical excitation densities $0.5n_0 < n_{\text{op}} < 2.6n_0$. This range is estimated for peak optical intensities 2.36 MW/cm$^2$ < $I_{\text{op}}$ < 11.8 MW/cm$^2$, using 2.6% absorbance [36] and 30% Fresnel coupling loss [37]. The inset of the Fig. 4 shows the underlying relationship of $I_{\text{op}}$ to $n_{\text{op}}$ and $D_s$. Agreement between experiment and theory validates the modification of the PSH dynamics by optical-doping.

Furthermore, the modified calculations can be applied to $S_z(0,y,t)$ and $S_z(x,0,t)$ in the presence of $|B_x|$ and $|B_y|$. The corresponding results are shown in Figs. 3(e) and (f). Calculations use previously determined parameters with the addition of $n_{\text{op}}$ and $D_{\text{s}}$ determined from the inset of Fig. (4). Now the S-shape behavior observed in the experiment is reproduced by the calculations, confirming that the initial excitation conditions are responsible for the spatiotemporal spin evolution and that light can tune/detune a PSH.

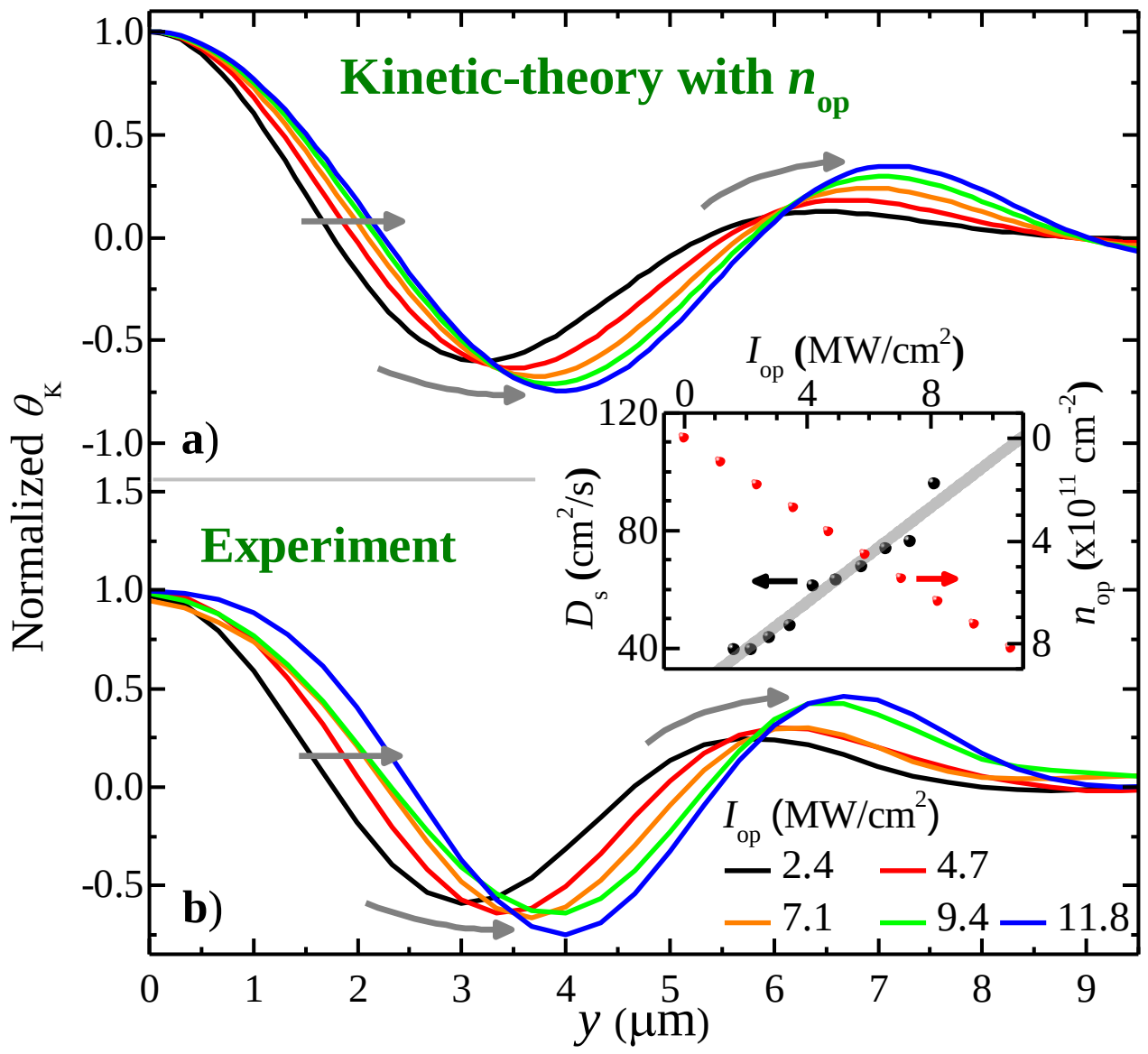

**Fig. 4** Modeled $S_z(x=0,y,t=750\text{ps})$ with SO parameters varying according to Eqs (6) and (7). (b) Experimental $S_z(x=0,y,t=0.75\text{ns})$ for a range of optical excitation densities. The inset relates the optical intensity $I_{op}$ to $n_{\text{op}}$ and $D_{\text{s}}$.

In conclusion, this work is the first experimental demonstration of spin-helix formation in II-VI 2DEGs. The evolution of the spin polarization of photoexcited carriers is found to be close to a persistent spin helix regime, where $\alpha$ and $\beta$ parameters are nearly equal. It is found that the optically induced carrier concentration affects both the Dresselhaus and Rashba SO coupling, and the diffusing spin distribution within the 2DEG. It is confirmed through a comparison of density-dependent experiments and an extended kinetic theory that the spin precession length is modified. This result establishes a dependence of spin helix on optical doping, which is important because low-dimension semiconductors can be specifically engineered, and the resulting spin helix is a model system for understanding SO coupling phenomena in topological and quantum matter.

**Acknowledgements**

We thank Dimitri Yakovlev and Evgeny Zhukov for useful discussions. We acknowledge the financial support by the Deutsche Forschungsgemeinschaft International Collaborative Research Centre TRR-160 (B1 and B3), the Government of the Russian Federation (contract #14.W03.3.0011 at Ioffe Institute), RFBR-DFG (project 15-52-12012), National Science Centre (Poland) Grants No. DEC- 2012/06/A/ST3/00247 and No. DEC-2014/14/M/ST3/00484, and by the Foundation for Polish Science through the IRA Programme co-financed by EU within SG OP. A.V.P also acknowledges support by Russian President Grant No. SP-2912.2016.5 and the Foundation “BASIS”.

F.P. and S.A. contributed equally to this work.